  \documentstyle[12pt]{article} 
  \topmargin - 0.5 cm 
\begin{document}

\thispagestyle{empty}

\noindent {\bf M. Daoud, M. Kibler}
\medskip

\noindent {\large\bf Fractional Supersymmetry\\ 
as a Superposition of Ordinary 
Supersymmetry\footnote{Paper written from a communication by M. Kibler 
at the ``X International Conference on Symmetry Methods in Physics'' 
(Yerevan, Armenia, 13 - 19 August 2003). To be published 
in Phys. Atom. Nuclei (Yadernaya fizika).}}

\bigskip

\bigskip

\noindent {A B S T R A C T}
\bigskip

\noindent
It is shown how to derive fractional supersymmetric 
quantum mechanics of order $k$ as a superposition of $k-1$ 
copies of ordinary supersymmetric quantum mechanics.

\newpage

\setcounter{page}{1}

\title {FRACTIONAL SUPERSYMMETRY AS A SUPERPOSITION OF 
          ORDINARY SUPERSYMMETRY}

\author{M. Daoud\footnote{Laboratoire de Physique 
de la Mati\`ere Condens\'ee, 
Facult\'e des Sciences, Universit\'e Ibn Zohr, 
BP 28/S, Agadir, Morocco},
        M. Kibler\footnote{Institut de Physique Nucl\'eaire de Lyon, 
IN2P3-CNRS et Universit\'e Claude Bernard, 43 Bd du 11 Novembre 1918, 
F-69622 Villeurbanne Cedex}}
\maketitle

\bigskip

\section{Introduction}
In recent years, fractional supersymmetry has been the subject 
of numerous works. Indeed, $k$-fractional supersymmetry is closely
connected to the notion of quantum algebra (deformation theory) and
to the concept of intermediate statistics (of anyons~\cite{Goldin} 
and $k$-fermions~\cite{DHK,chinois}) 
interpolating between Bose-Einstein statistics and Fermi-Dirac statistics.
Therefore, fractional supersymmetry constitutes a useful tool for 
dealing with anyonic statistics.
   
Fractional supersymmetric quantum mechanics of 
order $k$ can be considered as an extension of
ordinary supersymmetric quantum mechanics which
corresponds to $k=2$. An ordinary supersymmetric 
quantum-mechanical system may be generated from a 
doublet $(H,Q)_2$ of operators satisfying~\cite{Nicolai,Witten}
$$
Q^2 = 0, 
$$
$$
Q Q^{\dagger}  +  
Q^{\dagger} Q  =  H.
$$
The self-adjoint operator $H$ and the operator 
$Q$ act on a separable Hilbert space. The operator $H$ 
is referred to as the Hamiltonian and the
operator $Q$ as the supersymmetry operator of the ordinary 
supersymmetric quantum-mechanical system. The operator $Q$
gives rise to ${\cal N} = 2$ dependent supercharges 
$Q_- = Q$ and 
$Q_+ = Q^{\dagger}$ connected 
via Hermitean conjugation. They are nilpotent operators 
of order $k=2$ and commute with the Hamiltonian $H$.

The {\em ordinary} 
supersymmetric quantum-mechanical system $(H,Q)_2$ can be
extended to a {\em fractional} supersymmetric quantum-mechanical 
system $(H,Q)_k$ with $k \in {\bf N} \setminus \{ 0,1,2 \}$ as follows. 
The system $(H,Q)_k$ may be defined by~\cite{Rubakov,Khare}
$$
Q_- = Q,                             \quad 
Q_+ = Q^{\dagger}                    \quad 
(\Rightarrow \ Q_+ = Q_-^{\dagger}), \quad 
Q_{\pm}^k = 0, 
\eqno (1{\rm a})
$$
$$
Q_- ^{k-1} Q_+  +  Q_- ^{k-2} Q_+ Q_- 
                                      +   \cdots 
                                      +   Q_+ Q_- ^{k-1}
                                      =       Q_- ^{k-2} H,
\eqno (1{\rm b})
$$
$$
[H , Q_{\pm}] = 0, \quad H = H^{\dagger},
\eqno (1{\rm c})
$$
where the self-adjoint
operator $H$, the Hamiltonian of the system, and the ${\cal N} = 2$
supercharges $Q_-$ and 
             $Q_+$ act on a separable Hilbert space. Of course, 
the case $k=2$ corresponds to an ordinary supersymmetric 
quantum-mechanical system. 

In the present work, we study how it is possible to 
connect ordinary and $k$-fractional supersymmetric 
quantum-mechanical systems.

\section{The algebra $W_k$}
As an interesting question, we may ask: How
to construct a fractional supersymmetric quantum-mechanical 
system of order $k$ and, thus, fractional supersymmetric 
quantum mechanics of order $k$? This question can be answered 
through the definition of a generalized Weyl-Heisenberg algebra
$W_k$. We now define the generic algebra $W_k$ and shall see in 
the next section how a fractional supersymmetric quantum-mechanical 
system of order $k$ may be associated with a given algebra $W_k$. 

For $k$ given, with $k \in {\bf N} \setminus \{ 0,1 \}$, 
the algebra $W_k$ is generated by four linear operators 
$X_-$, $X_+$, $N$ and $K$. The operators $X_-$ and 
$X_+ = X_-^{\dagger}$ are shift operators connected 
via Hermitean conjugation. The operator $N$, called 
number operator, is self-adjoint. Finally, the 
operator $K$ is a $Z_k$-grading unitary operator. The 
generators $X_-$, $X_+$, $N$ and $K$ 
satisfy~\cite{DaoudKibler}
$$
 [X_- , X_+] = \sum_{s=0}^{k-1} f_s(N) \> \Pi_s, 
$$
$$
 [N , X_{-}] = {-} X_{-}, \quad (+{\rm h.c.}),
$$
$$
 [K , X_{-}]_{q} = 0,     \quad (+{\rm h.c.}),
$$
$$
 [K , N] = 0,             \quad K^k = 1.
$$
The functions $f_s : N \mapsto f_s(N)$ are such that 
$f_s(N)^{\dagger} = f_s(N)$, $[A , B]_{q}$ stands for $AB - qBA$,
and the operators $\Pi_{s}$ are defined by
$$
\Pi_{s} = \frac{1}{k}  \>  \sum_{t=0}^{k-1}  \>  q^{-st} \> K^t
$$
where 
$$
q = \exp \left( \frac{2 \pi {\rm i}}{k} \right)
$$
is a root of unity. To a given set $\{ f_s : s=0, 1, \cdots, k-1 \}$ 
corresponds one algebra $W_k$. 

The generalized Weyl-Heisenberg algebra $W_k$ 
covers numerous algebras describing exactly solvable 
one-dimensional systems. The particular system 
corresponding to a given set 
$\{ f_s : s = 0, 1, \cdots, k-1 \}$ yields, in a Schr\"odinger
picture, a particular dynamical system with a specific potential. Let 
us mention two interesting cases.
The case
$$
\forall s \in \{ 0, 1, \cdots, k-1 \} \ : \ 
f_s(N) = f_s \mbox{ independent of } N
$$
corresponds to systems with cyclic shape-invariant potentials 
(in the sense of Ref.~\cite{Sukhatme}) 
and the case
$$
\forall s \in \{ 0, 1, \cdots, k-1 \} \ : \ 
f_s(N) = a N + b,  \  (a,b) \in {\bf R}^2
$$
to systems with translational shape-invariant potentials 
(in the sense of Ref.~\cite{Junker}). For instance, 
the case $(a = 0, b > 0)$ 
corresponds to the harmonic oscillator potential, 
the case $(a < 0, b > 0)$ to the Morse potential and
the case $(a > 0, b > 0)$ to the P\"oschl-Teller potential. For these 
various potentials, the part of $W_k$ spanned by $X_-$, $X_+$ and $N$ 
can be identified with the ordinary Weyl-Heisenberg algebra 
for $(a = 0, b \not= 0)$,
                  with the su(2)   Lie algebra 
for $(a < 0, b > 0)$ and  
                  with the su(1,1) Lie algebra 
for $(a > 0, b > 0)$. 

\section{A $k$-frational system associated with $W_k$}
In order to associate a $k$-fractional supersymmetric 
quantum-mechanical 
system associated with a given generalized Weyl-Heisenberg 
algebra $W_k$, we must define a supersymmetry operator $Q$ 
and an Hamiltonian $H$. The supersymmetry operator $Q$ is 
defined by
$$
Q           \equiv Q_- = X_- (1 - \Pi_{1}) \Leftrightarrow 
Q^{\dagger} \equiv Q_+ = X_+ (1 - \Pi_{0}). 
$$
Then, the Hamiltonian $H$ associated with $W_k$ can be deduced 
from Eq.~(1b). This yields 
$$
H = (k-1) X_+ X_- -
\sum_{s=3}^k 
\sum_{t=2}^{s-1} (t-1) \> f_t(N-s+t) \> \Pi_s 
$$
$$
- 
\sum_{s=1}^{k-1} 
\sum_{t=s}^{k-1} (t-k) \> f_t(N-s+t) \> \Pi_s.
$$ 
(Note that the summation from $s = k-2$ to $s = k$ appearing
in some previous works by the authors~\cite{DaoudKibler} should
be replaced by a summation from $s = 3$ to $s = k$.) It can be 
checked that $H$ is self-adjoint and commutes 
with $Q_-$ and $Q_+$. As a conclusion, 
the doublet $(H, Q)_k$ associated to $W_k$
satisfies Eq.~(1) and thus defines a $k$-fractional 
supersymmetric quantum-mechanical system.

\section{Connection between fractional supersymmetry and ordinary supersymmetry}
In order to establish a connection between {\em fractional}
supersymmetric quantum mechanics of order $k$ and {\em ordinary} 
supersymmetric quantum mechanics (corresponding to $k = 2$),
it is necessary to construct sub-systems from the doublet $(H, Q)_k$ 
that correspond to ordinary supersymmetric quantum-mechanical systems.
This may be achieved in the following way~\cite{PLA}. The general 
Hamiltonian $H$ can be 
rewritten as 
$$
H = \sum_{s=1}^{k} H_{s} \> \Pi_{s} 
$$
where 
$$
H_s \equiv H_s(N) 
    = (k-1) X_+ X_-  -  \sum_{t=2}^{k-1} (t-1) \> f_t(N-s+t)
$$    
$$    
+ 
(k-1)        \sum_{t=s}^{k-1}          f_t(N-s+t). 
$$
It can be shown that the operators 
$H_k \equiv H_0, H_{k-1}, \cdots, H_1$ 
turn out to be isospectral operators. 
It is possible to factorize $H_s$ as~\cite{PLA} 
$$
H_s = X(s)_+ \> X(s)_-.
$$
Let us now define: (i) the two (supercharge) operators
$$
q(s)_- = X(s)_- \> \Pi_s, \quad 
q(s)_+ = X(s)_+ \> \Pi_{s-1}
$$ 
and (ii) the (Hamiltonian) operator 
$$
h(s) = X(s)_- \> X(s)_+ \> \Pi_{s-1}   +   X(s)_+ \> X(s)_- \> \Pi_s.
$$
It is then a simple matter of calculation to prove that 
$h(s)$ is self-adjoint and that
$$
q(s)_+ = q(s)_-^{\dagger},             \quad
q(s)_{\pm}^2 = 0,                      \quad 
h(s) = \{ q(s)_- , q(s)_+ \},          \quad 
[ h(s) , q(s)_{\pm} ] = 0. 
$$
Consequently, the doublet $(h(s), q(s))_2$, with
$q(s) \equiv q(s)_-$, 
satisfies Eq.~(1) with $k=2$ and thus 
defines an ordinary supersymmetric 
quantum-mechanical system (corresponding to $k=2$). 

The Hamiltonian $h(s)$ is amenable to a form more 
appropriate for discussing the link between ordinary 
supersymmetry and fractional supersymmetry. Indeed, we can show that 
$$ 
            X(s)_- \> X(s)_+ = H_s(N+1).
$$
Then, we can obtain the important relation 
$$
h(s) = H_{s-1} \> \Pi_{s-1} + H_{s} \> \Pi_{s}
$$
to be compared with the expansion of $H$ in terms 
of supersymmetric partners $H_s$. 

As a result, the system $(H,Q)_k$, corresponding to $k$-fractional
supersymmetry, can be described in terms of $k-1$ sub-systems $(h(s),q(s))_2$, 
corresponding to ordinary supersymmetry. The Hamiltonian $h(s)$ is given 
as a sum involving the supersymmetric partners 
$H_{s-1}$ and $H_s$. Since the 
supercharges $q(s)_{\pm}$ commute with the Hamiltonian $h(s)$, 
it follows that 
$$
H_{s-1} X(s)_- = X(s)_- H_{s  }, \quad 
H_{s  } X(s)_+ = X(s)_+ H_{s-1}.
$$
As a consequence, the operators 
          $X(s)_+$ 
      and $X(s)_-$ render 
possible to pass from the spectrum of $H_{s-1}$ and $H_{s  }$
                        to the one of $H_{s}  $ and $H_{s-1}$, 
respectively. This result is quite familiar for ordinary 
supersymmetric quantum mechanics (corresponding to $s=2$).

For $k=2$, the operator $h(1)$ is nothing but the total Hamiltonian $H$ 
corresponding to ordinary supersymmetric quantum mechanics. For
arbitrary $k$, the other operators $h(s)$ are simple replicas (except for the
ground state of $h(s)$) of $h(1)$. In 
this sense, fractional  supersymmetric 
quantum mechanics of order $k$ can be 
considered as a set of $ k-1 $ replicas 
of ordinary supersymmetric quantum mechanics
corresponding to $k=2$ and 
typically described by $(h(s),q(s))_2$. As a further argument, 
it is to be emphasized that 
$$
H = q(2)_- \>
    q(2)_+ + \sum_{s=2}^{k} 
    q(s)_+ \> 
    q(s)_-
$$
which can be identified with $h(2)$ for $k=2$.

\section{Conclusions}

Starting from a $Z_k$-graded algebra $W_k$,
characacterized by a set $ \{ f_s : s = 0, 1, \cdots, k-1 \} $
of structure functions,
it was shown how to associate a $k$-fractional supersymmetric 
quantum-mechanical system $(H,Q)_k$ 
characterized by an Hamiltonian $H$
and a supercharge $Q$. 

The Hamiltonian $H$ for the system $(H,Q)_k$
was developed as a superposition of $k$ isospectral 
supersymmetric partners $H_k, H_{k-1}, \cdots, H_{1}$. 
It was proved that the system $(H,Q)_k$ can be described 
in terms of $k-1$ sub-systems $(h(s),q(s))_2$ which are 
ordinary supersymmetric quantum-mechanical systems.

\end{document}